\documentclass[9pt,twocolumn,twoside]{pnas-new}

\templatetype{pnasresearcharticle}

\title{Laboratory evidence of dynamo amplification of magnetic fields in a turbulent plasma}

\author[a,b]{P. Tzeferacos}
\author[a]{A. Rigby} 
\author[a]{A. Bott}
\author[a]{A. R. Bell}
\author[c,d]{R. Bingham}
\author[e]{A. Casner}
\author[b]{F. Cattaneo}
\author[f,g]{E. M. Churazov}
\author[h]{J. Emig}
\author[i]{F. Fiuza}
\author[j]{C. B. Forest}
\author[k]{J. Foster}
\author[b]{C. Graziani}
\author[l]{J. Katz}
\author[m]{M. Koenig}
\author[n]{C.-K. Li}
\author[a]{J. Meinecke}
\author[n]{R. Petrasso}
\author[h]{H.-S. Park}
\author[h]{B. A. Remington}
\author[h]{J. S. Ross}
\author[o]{D. Ryu}
\author[h]{D. Ryutov}
\author[a]{T. G. White}
\author[p]{B. Reville}
\author[q]{F. Miniati}
\author[a]{A. A. Schekochihin}
\author[b]{D. Q. Lamb}
\author[l]{D. H. Froula}
\author[a,b,1]{G. Gregori} 

\affil[a]{Department of Physics, University of Oxford, Parks Road, Oxford OX1 3PU, UK}
\affil[b]{Department of Astronomy and Astrophysics, University of Chicago, 5640 S. Ellis Ave, Chicago, IL 60637, USA}
\affil[c]{Rutherford Appleton Laboratory, Chilton, Didcot OX11 0QX, UK}
\affil[d]{Department of Physics, University of Strathclyde, Glasgow G4 0NG, UK}
\affil[e]{CEA, DAM, DIF, F-91297 Arpajon, France}
\affil[f]{Max Planck Institute for Astrophysics, Karl-Schwarzschild-Strasse 1, D-85741 Garching, Germany}
\affil[g]{Space Research Institute (IKI), Profsouznaya 84/32, Moscow, 117997, Russia}
\affil[h]{Lawrence Livermore National Laboratory, Livermore, CA 94550, USA}
\affil[i]{SLAC National Accelerator Laboratory, 2575 Sand Hill Road, Menlo Park, CA 94025, USA}
\affil[j]{Physics Department, University of Wisconsin-Madison, 1150 University Avenue Madison, WI 53706, USA}
\affil[k]{AWE, Aldermaston, Reading, West Berkshire, RG7 4PR, UK}
\affil[l]{Laboratory for Laser Energetics, University of Rochester, 250 E. River Rd, Rochester, NY 14623, USA}
\affil[m]{Laboratoire pour l'Utilisation de Lasers Intenses, UMR7605, CNRS CEA, Universit\'e Paris VI Ecole Polytechnique, 91128 Palaiseau Cedex, France}
\affil[n]{Massachusetts Institute of Technology, Cambridge, Massachusetts 02139, USA}
\affil[o]{Department of Physics, UNIST, Ulsan 689-798, Korea}
\affil[p]{School of Mathematics and Physics, Queens University Belfast, Belfast BT7 1NN, UK}
\affil[q]{Department of Physics, ETH Z\"urich, Wolfgang-Pauli-Strasse 27, CH-8093 Z\"urich, Switzerland}

\leadauthor{Tzeferacos} 

\authorcontributions{This project was conceived by G. Gregori, D.Q. Lamb, P. Tzeferacos, B. Reville, and A.A. Schekochihin. The experimental team was led by G. Gregori and the simulation work by P. Tzeferacos. The Thomson scattering polarimetry diagnostics at the Omega laser facility was designed by D.H. Froula and J. Katz. 
All authors have provided contributions to either data analysis, MHD simulations or theoretical support.}
\authordeclaration{The authors declare no conflict of interest.}
\correspondingauthor{\textsuperscript{1}To whom correspondence should be addressed. E-mail: gianluca.gregori@physics.ox.ac.uk}

\keywords{Magnetic fields $|$ MHD Turbulence $|$ Dynamo} 

\begin{abstract}
Magnetic fields are ubiquitous in the Universe. Extragalactic disks, halos and clusters have consistently been shown, via diffuse radio-synchrotron
emission and Faraday rotation
measurements, to exhibit magnetic field strengths ranging from a few nG to tens of $\mu$G \cite{Carilli2002}. The energy density of these
fields is typically comparable to the energy density of the fluid motions of the plasma in which they are embedded, making
magnetic fields essential players in the dynamics of the luminous matter. 
The standard theoretical model for the origin of these strong magnetic fields is through the amplification of tiny seed 
fields via turbulent dynamo to the level consistent with current observations \cite{Subramanian2006a,Ryu2008,Miniati2015,Zweibel1997,Schekochihin2006}. 
Here we demonstrate, using laser-produced colliding plasma flows, that turbulence is indeed capable of rapidly amplifying seed fields to
near equipartition with the turbulent fluid motions. These results support the notion that 
turbulent dynamo is a viable mechanism responsible for the observed present-day magnetization of the Universe.
\end{abstract}

\dates{This manuscript was compiled on \today}

\begin{document}

\verticaladjustment{-2pt}

\maketitle
\thispagestyle{firststyle}
\ifthenelse{\boolean{shortarticle}}{\ifthenelse{\boolean{singlecolumn}}
{\abscontentformatted}{\abscontent}}{}

\noindent
That turbulence is of central importance in the generation and evolution of magnetic fields in the Universe is essentially
without doubt \cite{Zweibel1997}. Plasma turbulence can be found in myriads of astrophysical objects, 
where it is excited by a range of processes: cluster mergers, supernovae explosions, stellar outflows,
etc. \cite{Subramanian2006a,Schekochihin2006,Ryu2008}. 
If a turbulent plasma is threaded by a weak magnetic field, the stochastic motions of the fluid will stretch
and fold this field, amplifying it until it becomes dynamically significant \cite{Batchelor1950,Biermann1951}.
According to the current standard picture, the amplification can be summarized in two basic steps \cite{Schekochihin2004,Beresnyak2012}.
First, when the initial field is small, the magnetic energy grows exponentially (kinematic phase). This phase
terminates when the magnetic energy reaches approximate equipartition with the kinetic energy at the dissipation scale.
Beyond this point, the magnetic energy continues to grow linearly in time (non-linear phase), until, after
roughly one outer-scale eddy turnover time, it saturates at a fraction of the total kinetic energy of
the fluid motions \cite{Haugen2004,Schekochihin2004}. This is what is referred to as the turbulent dynamo
mechanism for magnetic field amplification.

The seed fields that the dynamo amplifies can be produced by a variety of different physical processes.
In many astrophysical environments where the plasma is initially unmagnetized, and most certainly at the
time when proto-galaxies were forming, baroclinic generation of magnetic fields due to misaligned density and temperature gradients -- the
Biermann battery mechanism -- can provide initial seeds \cite{Kulsrud1997}.
The same starting fields also occur in laser produced plasmas \cite{Stamper1971,Gregori2012}.

While the theoretical expectations that turbulent dynamo must operate go back more than half a
century \cite{Batchelor1950,Biermann1951} and the first direct numerical confirmation of this
effect was achieved 35 years ago \cite{Meneguzzi1981HelicalDynamos}, 
detecting turbulent dynamo amplification in the laboratory has remained elusive. This is primarily because of the difficulty of
achieving experimentally magnetic Reynolds numbers (${\rm Rm} = u_L L/\mu$, where $u_L$ is the flow
velocity at the outer scale $L$, and $\mu$ is the magnetic diffusivity) above the critical threshold
of a few hundred required for dynamo \cite{Schekochihin2007}.
Such a demonstration would not only establish experimentally the soundness of the existing theoretical
and numerical expectations for one of the most fundamental physical
processes in astrophysics \cite{Kulsrud1995ImportantAstrophysics}, but also provide a
platform \cite{Gregori2015} to investigate other fundamental processes that require the
a turbulent magnetized plasma, such as particle acceleration and reconnection.

To date, experimental investigation of magnetic-field amplification has primarily been carried out
in liquid-metal experiments \cite{Monchaux2007}, where the dynamo that was achieved was of a mean-field
type and depended on a particular fluid flow, rather than a purely turbulent effect leading to a stochastic field.
More recent work has focused on laser-driven plasmas \cite{Meinecke2014a,Meinecke2015a,Gregori2015},
but studying a regime that is a precursor to dynamo, because of the modest magnetic Reynolds numbers that could be achieved. 

In the experiment described here, we reach magnetic Reynolds numbers above the expected dynamo threshold. 
These experiments were performed at the Omega laser facility at the Laboratory for Laser Energetics of the
University of Rochester \cite{Boehly1997} using a combined platform that builds on our previous work on smaller
laser facilities \cite{Gregori2012,Meinecke2014a,Meinecke2015a}. Laser ablation of a chlorine-doped plastic
foil launches a plasma flow from its rear surface. The plasma then passes through a solid grid and collides
with an opposite moving flow, produced in the same manner. In order to increase the destabilization of the
motions as the flows collide, the two grids have hole patterns that are shifted with respect to each other.
Further details on the experimental setup are given in Figure 1. A set of diagnostics has been fielded to
measure the properties of the flow, its turbulence and the magnetic field generated by it (see Figures 2 and 3).

Extensive two-dimensional and three-dimensional simulations done prior to the experiments using the
radiation-magnetohydrodynamics code FLASH informed their design (see Supplementary Information),
including the details of the targets and the grids, and the timing of the
diagnostics \cite{Tzeferacos2015a,Tzeferacos2016}.

\begin{figure}
\centering
\includegraphics[width=1\linewidth]{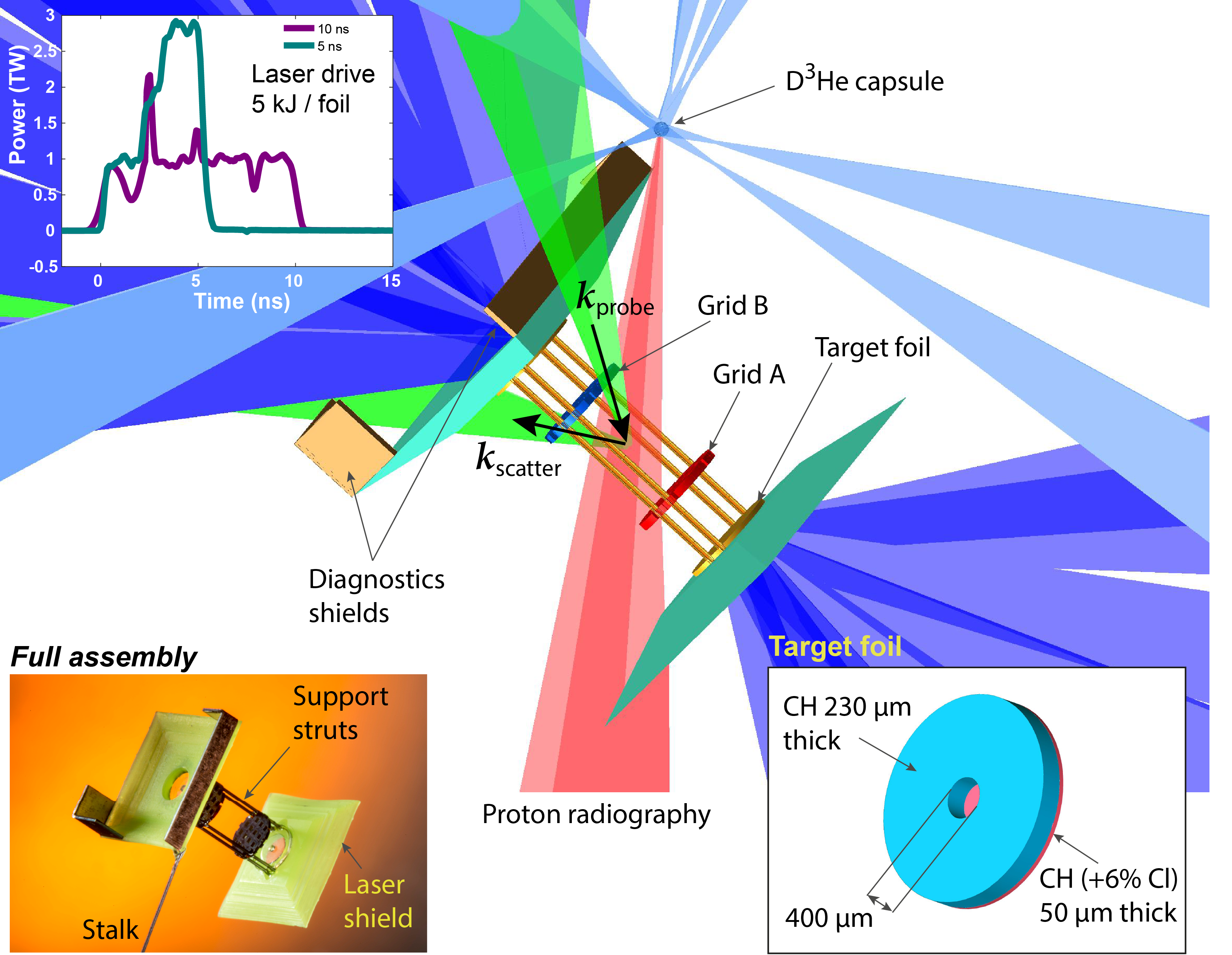}
\caption{{\bf Laser, target and diagnostics configuration.}  Two CH foils doped with 6\% chlorine in
atomic number (details given in the lower right panel), separated by 8 mm, are each illuminated by
ten 500 J, 1 ns pulse length, frequency tripled (351 nm 
wavelength) laser beams with 800 $\mu$m spot diameter. The beams are stacked in time to achieve the
two pulse profiles shown in the upper left panel. An additional set of 17 beams, all fired simultaneously,
are used to implode a 420 $\mu$m diameter capsule consisting of a 2 $\mu$m  
thick $\rm SiO_2$  shell filled with $\rm D_2$ gas at 6 atm and $\rm {}^3He$ at 12 atm. The implosion
produces mono-energetic protons at 3.3 MeV and 15 MeV with $\sim$40 $\mu$m diameter source size, which
traverse the plasma and are then collected by a CR-39 nuclear track detector with a total magnification 
factor of 28. 
The plasma expansion towards the center of the target is perturbed by the presence of two grids,
placed 4 mm apart, with a 300 $\mu$m hole width and 300 $\mu$m hole spacing. Grid A has the central
hole aligned on the center axis connecting the two foils, while grid B has the hole pattern shifted
so that the central axis crosses the middle point between two holes.
Thomson scattering uses a 30 J, 1 ns, frequency doubled (wavelength $\lambda = 526.5$ nm) laser beam
to probe the plasma on the axis of the flow, 400 $\mu$m from the center and in a 50 $\mu$m focal spot,
towards grid B. The scattered light is collected with 63$\rm ^o$
scattering angle and the geometry is such that the scattering wavenumber 
${\bf k} = {\bf k}_{\rm scatter}-{\bf k}_{\rm probe}$, where $|{\bf k}_{\rm scatter}|\approx|{\bf k}_{\rm probe}|=2\pi / \lambda$, is parallel to the axis of the flow.}
\label{Fig1}
\end{figure}

X-ray emission can be used to characterize the interaction of the colliding jets and assess properties
of the resulting plasma inhomogeneities. The presence of a small amount of chlorine in the plasma enhances the
emission in the soft wavelength region ($< 2$ keV). Soft X-ray images taken at $t=35$ ns from the start of
the laser drive, which is after the flows collide, indicate a broad non-uniform spatial distribution of the
emission over a region more than 1 mm across. 
In order to characterize the state of this interaction region, power spectra of the X-ray intensity fluctuations
were extracted from the experimental data using a two-dimensional fast Fourier transform (see Figure 2).
Under the assumption of isotropic statistics, fluctuations in the detected X-ray intensity are directly related
to density fluctuations (see discussion in Supplementary Information). The power spectrum of the density
fluctuations extracted from the X-ray data is consistent with a Kolmogorov power law ($k^{-5/3}$ scaling).
Experimental data from other diagnostics indicates that the plasma motions are mainly subsonic
(Mach number $\lesssim$1 at the outer scale); as a result, density fluctuations injected at large
scales behave as a passive scalar and the spectra of the density and velocity fluctuations should be the
same \cite{Zhuravleva2014}. We conclude that the X-ray emission 
supports the notion that turbulent motions are present in the interaction region.
This is also confirmed by FLASH simulations~\cite{Tzeferacos2016}, which predict subsonic
motions of the plasma following the jet collision. Furthermore, the power spectrum of
density and velocity fluctuations can be calculated directly from FLASH, and the results are consistent the same power
law scaling for both (Figures 2c and 2d). 

\begin{figure}
\begin{center}
\includegraphics[width=1.05\linewidth]{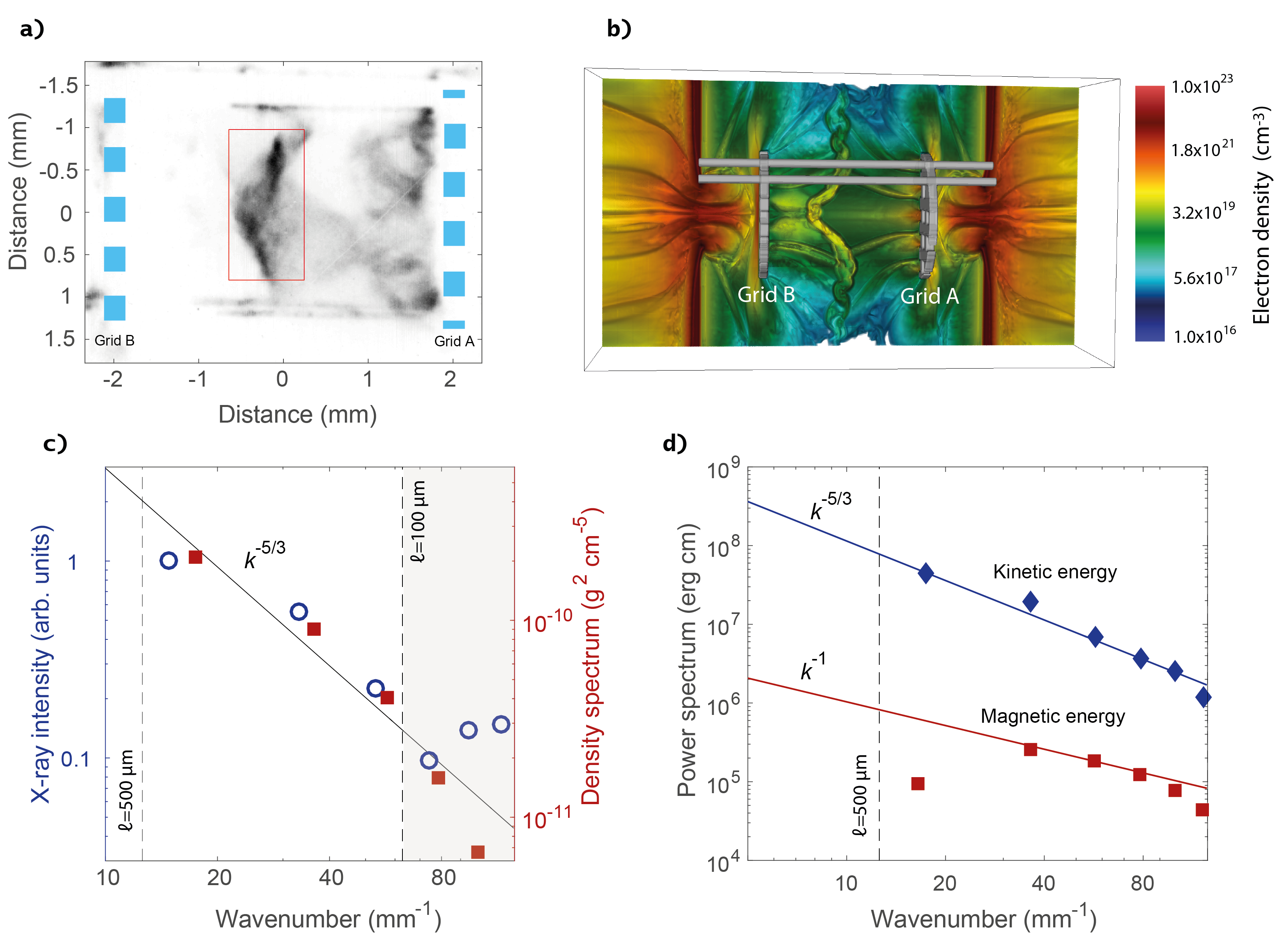}	
\caption{{\bf Characterization of the plasma turbulence.}  (a) X-ray pinhole image of the colliding
flows at $t=35$ ns after the laser drive, using the 5 ns pulse profile.
The image was recorded onto a framing camera with $\sim$1 ns gate width and filtered
with 0.5 $\mu$m $\rm C_2H_4$ and 0.15 $\mu$m Al. The pinhole diameter is 50 $\mu$m.  
(b) Rendering of the electron density from three-dimensional FLASH simulations at $t=35$ ns.
(c) The open blue circles give the power spectrum of the X-ray emission from the collision
region, defined by the rectangular region shown in panel (a). The power spectrum has been
filtered to remove high-frequency noise and edge effects. Details of this procedure are
given in the Supplementary Information. The spectrum of the density fluctuations, as obtained
from FLASH simulations in the jet collision region, is shown with red squares.
(d) Blue diamonds: power spectrum of the kinetic energy from FLASH simulations. Red squares:
power spectrum of magnetic energy from FLASH simulations.}
\end{center}
\end{figure}

The Thomson scattering diagnostic (see Figure 1 and Supplementary Information) 
allows us to measure simultaneously three different velocities associated with the flow \cite{Evans1969}. 
First, the bulk plasma-flow velocity -- composed of a mean flow velocity $U$ and outer-scale
turbulent velocity $u_L$ -- is obtained from the measurements of red shifts (in frequency) of
the scattered light resulting from the bulk plasma moving towards grid A. Secondly, the separation
of the ion-acoustic waves is an accurate measure of the sound speed and thus of the electron
temperature, $T_e$. Thirdly, the FLASH prediction of equal ion and electron temperatures allows
us to infer from the broadening of the ion-acoustic features the turbulent velocity $u_{\ell}$ on
the scale $\ell \sim 50$ $\mu$m (the Thomson scattering focal spot) \cite{Inogamov2003,Meinecke2015a}.

\begin{figure}
\begin{center}
\includegraphics[width=0.78\linewidth]{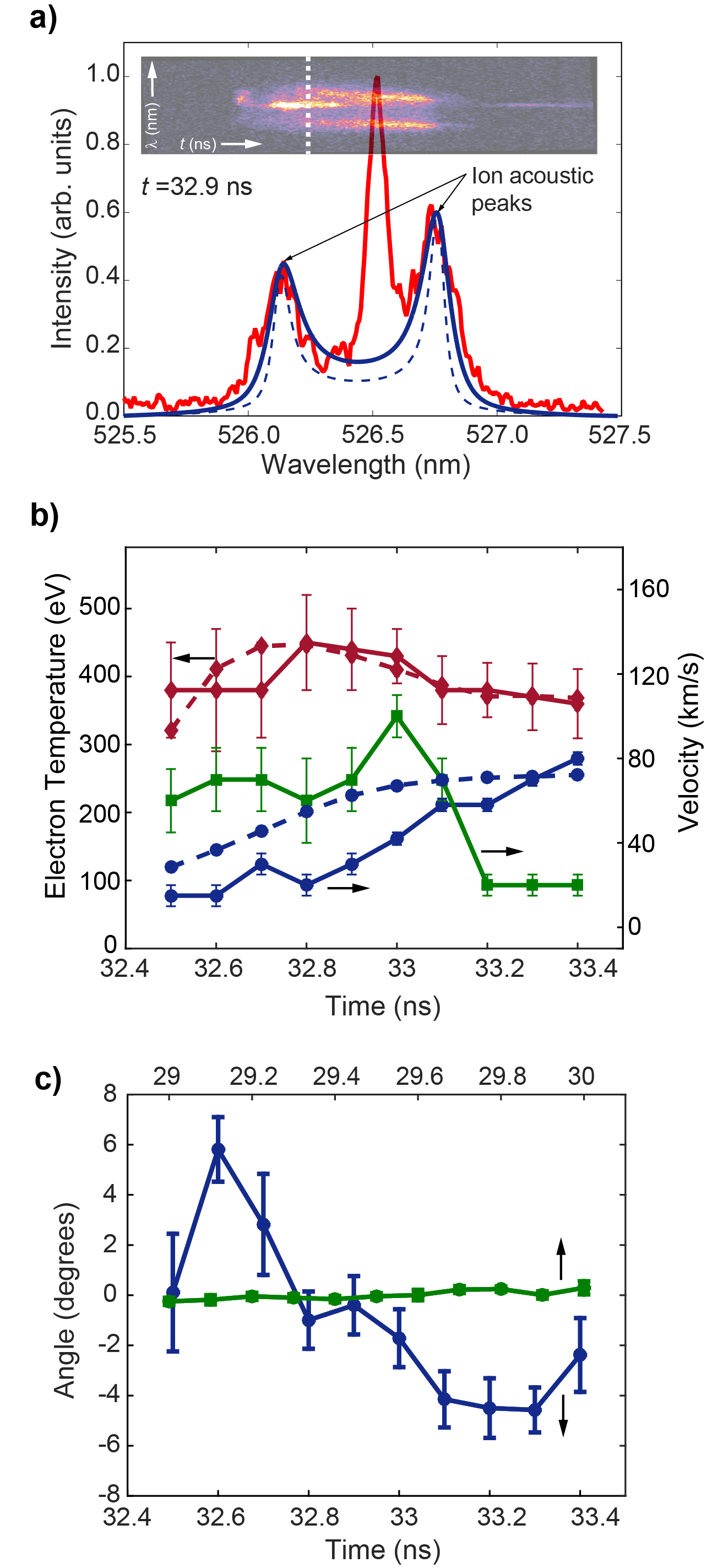}
\caption{{\bf Thomson scattering measurements}. 
Electron temperatures and flow velocities are obtained by fitting the experimental data with the
frequency dependent Thomson scattering cross section \cite{Evans1969}. In the fitting procedure we
assumed an electron density of $\lesssim 10^{20}$ cm$^{-3}$  (as predicted by FLASH simulations). At
these electron densities, the frequency distribution of the scattered light does not depend on the electron density,
which only provides an overall normalization factor.
(a) Thomson scattering data (red solid line) at $t=32.9$ ns obtained from a target driven with the 5 ns pulse profile.
The blue dashed line corresponds to a plasma in thermodynamic equilibrium (assuming equal electron and ion temperatures).
The central peak is due to stray light at the probe laser wavelength (and it is used to determine the instrumental resolution of the spectrometer).  
The blue solid line corresponds to the case in which additional broadening due to turbulence is included in the fitting procedure.
The inset in the top panel shows the time-streaked image of the Thomson scattered light. The resolution of the streak
camera is $\sim$ 50 ps and the Thomson scattering signal is fitted every 100 ps.
(b) Flow velocity towards grid A (full blue circles), turbulent velocity (full green squares) and
electron temperature (full red diamonds) as measured by Thomson scattering for the case of a target driven
with 5 ns laser profile. FLASH simulation results for the electron temperature and flow velocity in the probe
volume are also reported in dashed lines. The error bars are estimated from the $\chi^2$ fit of the data.  
(c) Estimated Faraday rotation data from the Thomson scattering data. This was done by separating the scattered
light into two orthogonal polarizations (see Supplementary Information). The blue line corresponds to the same
conditions as (b) above. The green line was obtained for the case of a single grid only, when the magnetic field is expected to be significantly smaller
(see Fig.~9 in the Supplementary Information).}
\end{center}
\end{figure}

Based on these measurements, we find the following. Before the collision, the two plasma flows move towards each other with axial mean velocity  $U \lesssim 200$ km/s
in the laboratory rest frame, and have an electron temperature $T_e \approx 250$ eV (see Figure 5 in the
Supplementary Information). After the collision, the axial flow slows down to $20$-$40$ km/s, with motions
being converted into transverse components. The electron temperature increases considerably, reaching
$T_e \approx 450$ eV (Figure 3). The measured time-averaged (RMS) turbulent velocity at scale $\ell$ is
$u_{\ell} \sim 55$ km/s. If $u_\ell$ has Kolmogorov scaling, the turbulent velocity at the outer
scale must therefore be $u_L \sim u_\ell (L/\ell)^{1/3} \approx 100$ km/s. Electron density estimates can be obtained from the measured total
intensity of the Thomson scattered radiation, to give a value $n_e \approx 10^{20} \, {\rm cm^{-3}}$,
which is also consistent with values predicted by FLASH simulations \cite{Tzeferacos2016}. As shown in the
Supplementary Information, a plasma with these parameters can be well described as being in the
magneto-hydrodynamic (MHD) regime.

For an MHD-type plasma, we can estimate the characteristic fluid and magnetic Reynolds numbers attained in our experiment. We
find ${\rm Re} = u_L L/\nu \sim 600$ ($\nu$ is the viscosity), and ${\rm Rm} \sim 700$,
using $L\sim 600$ $\mu$m, the characteristic driving scale determined by the average separation
between grid openings.  We have thus achieved conditions where ${\rm Rm}$ is comfortably larger
than the expected critical magnetic Reynolds number required for turbulent dynamo \cite{Schekochihin2007}.
The experiment also lies in the regime where the magnetic Prandtl number is
${\rm Pm}={\rm Rm}/{\rm Re} \lesssim 1$.

Magnetic fields were inferred using both Faraday rotation (Figure 3) and proton radiography (Figure 4).
The rotation of the polarization angle of Thomson scattered light provides a measure of the variation
of the longitudinal component of the magnetic field integrated along the beam path, weighted by the electron
density. 
Assuming a random field with correlation length $\ell_B$, we estimate
$B_{\parallel,rms} \approx 120 \, (\Delta \theta/3^{\circ})  (n_e /10^{20} \, {\rm cm^{-3}})^{-1} (\ell_n \ell_B / 0.2 \, {\rm mm^2})^{-1/2}$ kG, 
where $B_{\parallel,rms}$ is the root mean square (RMS) value of the magnetic field component parallel to the probe beam,
$\Delta \theta$ is the rotation angle, and $\ell_n \sim L \approx 0.6$ mm is the scale length of the electron density along the
line of sight.
Estimating $\ell_B$  is more challenging, but as a reasonable estimate we can take the size of the grid aperture
($\ell_B \sim 300$ $\mu$m). 

\begin{figure}[h]
\begin{center}
\includegraphics[width=0.7\linewidth]{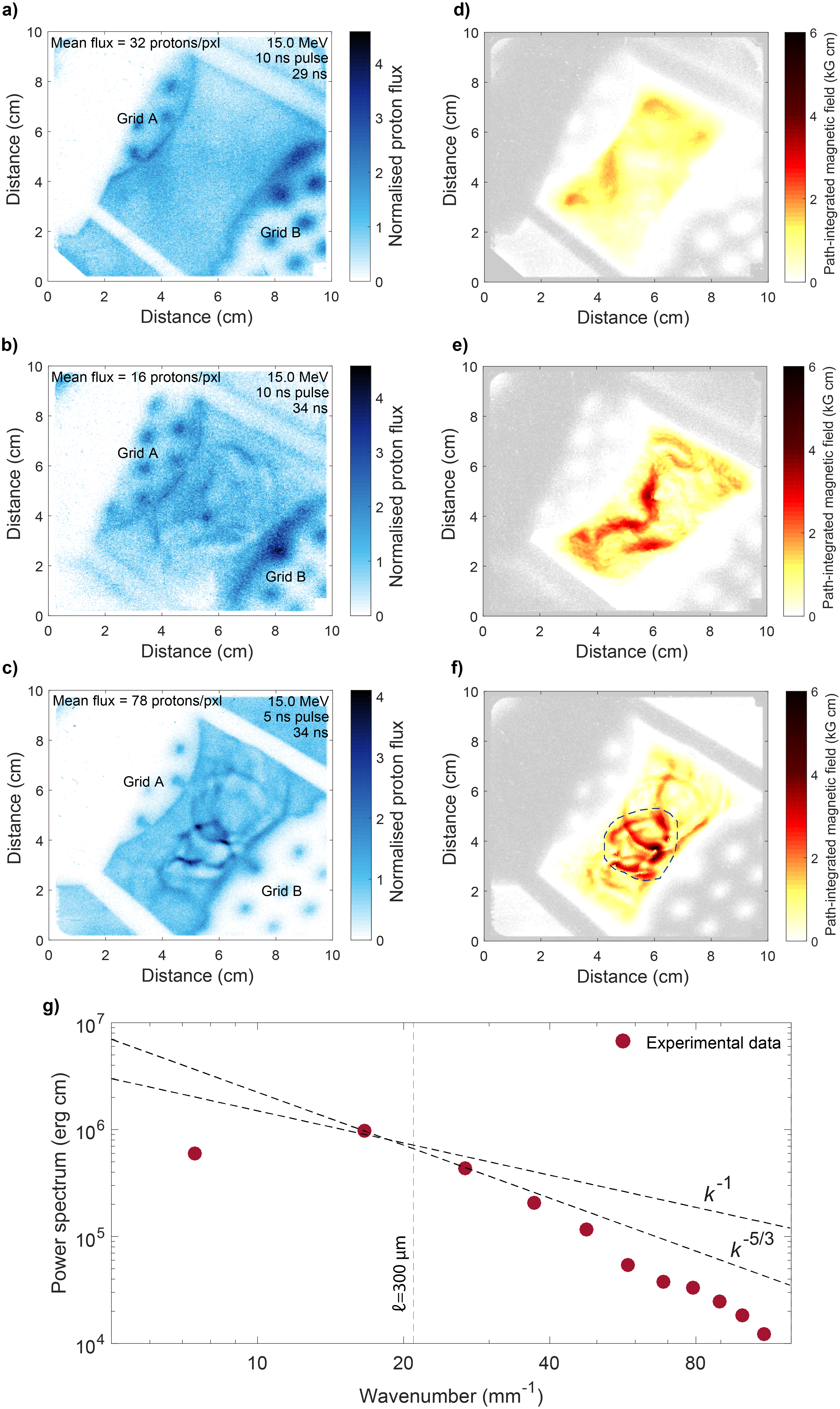}
\caption{{\bf Proton radiography.}  
(a) Normalized number of 15 MeV protons detected on a CR-39 plate. The normalization is such that
unity corresponds to the mean number of protons per pixel on the detector.
The D$^3$He capsule was imploded at $t=29$ ns. Fusion reactions occur 0.6 ns after the start of the
implosion and the protons are emitted isotropically within a short burst, of $\sim$150 ps
duration \cite{Li2006}. The flight time of the protons to the plasma is 0.1 ns. The chlorinated
plastic foils were driven with a 10 ns long pulse shape (see Figure 1). X-ray data and FLASH simulations
indicate that the plasma flows are close to collision by 29 ns (see also Figure 11 in the Supplementary
Information). Thus, this proton image can provide an estimate of the initial seed fields.  
(b) Same as (a), but with the deuterium-tritium capsule imploded at $t=34$ ns. The development of structures shows the development of fields in the interaction region. 
(c) Same as (b), but with the chlorinated plastic foils driven with the 5 ns long pulse, which gives higher flow velocities, and hence higher magnetic Reynolds numbers.
(d) Reconstruction of magnetic fields for case (a). 
(e) Reconstruction of magnetic fields for case (b). 
(f) Reconstruction of magnetic fields for case (c).
(g) Power spectrum of the magnetic energy from the reconstructed magnetic field from experimental data (the region bound by a dashed line in panel 4f).}
\end{center}
\end{figure}

The appearance of strong, sharp features
in proton radiographs provides independent evidence for large magnetic fields \cite{Li2006,Kugland2012a}.
Such structures are the 
result of initially divergent proton rays produced by an imploding D$^3$He capsule being focused by magnetic forces as they traverse the plasma.
This leads, at the detector plane, to localized regions where the proton counts greatly exceed their average value and other regions where they are strongly depleted. 
The detailed spatial structure of the path-integrated magnetic field can be reconstructed from the experimental images assuming that the protons undergo small deflections
as they pass through the plasma, and that the paths of neighboring protons do not cross before reaching the screen. Assuming isotropic statistics, this is sufficient information to calculate the power spectrum of the magnetic energy $E_B\!\left(k\right)$, and then the RMS magnetic field strength $B_{rms}$ of fluctuating fields via $B_{rms}^2 = 8 \pi \int \mathrm{d}k \, E_B\!\left(k\right)$ (see Ref.~\cite{Graziani2016} and Supplementary Information).
Figure 4 shows that the magnetic field during the early phases of the collision (see panel a) is small, as no strong flux features appear in the radiographic image. The corresponding reconstructed RMS magnetic field strength $B_{rms}$ obtained from Figure 4d gives $B_{rms} \lesssim  4\, $kG.
Magnetic fields before the collision, and in the absence of any strong turbulence, are presumably
Biermann battery fields produced at the laser spots and then advected by the flow, as indeed is confirmed by FLASH simulations.
In contrast, Figures 4b and 4c -- corresponding to a later stage of the turbulent plasma's lifetime for
the 10 ns and 5 ns pulse shapes, respectively -- do indeed show strong features, indicative of increased fields
strengths and altered morphology. The reconstruction algorithm can also be applied to the images in Figures
4b and 4c; in the latter case, we obtain $B_{rms} \approx 100$ kG (see Supplementary Information). This is consistent with our previous
estimates based on Faraday rotation. We claim that this increase of the magnetic field during the collision
cannot be simply explained by the compression of the field lines due to the formation of shocks
(this would only account for a factor of two increase at most), nor by
further generation by Biermann battery as the temperature gradients are not strong enough. This view is supported by
FLASH simulations (see Supplementary Information).

In Figure 4g we show the spectrum of the magnetic energy, $E_B\!\left(k\right)$, calculated from the
reconstructed path-integrated magnetic field. This is the spectrum on which the estimate of $B_{rms}$ is based.
The peak of
this spectrum occurs at a wavenumber consistent with the claim that energetically
dominant magnetic structures have a size $\ell_B \sim 300$ $\mu$m. The slope of the spectrum steepens at small wavelengths ($\lesssim 100$ $\mu$m),
which is likely due to diffusion of the imaging beam caused by small-scale magnetic fields, and the underestimation of the magnetic energy by the
reconstruction algorithm in the presence of small-scale caustics (see Supplementary Information for a discussion of such effects).
In the FLASH simulations the magnetic field spectrum appears to have a
$\sim k^{-1}$ power-law dependence, as shown in Figure 2d, consistent with the spectra of
tangled fields near and above the dynamo threshold found in Ref.~\cite{Schekochihin2007}.

Our experiment thus indicates that, as the two plasma flows collide, a strongly turbulent plasma, with
magnetic Reynolds number above the threshold for dynamo action, is generated. The magnetic field grows
from an initial value $B_{rms} \lesssim 4$ kG to $\sim 100$-$120$ kG.
We assume this to be near the saturated value because the Faraday rotation measurement begins over $2$ ns (comparable to dynamical times)
before the proton imaging diagnostic,
and we infer similar magnetic field strengths from both. Note that the expected timescale for saturation
to be reached is of the order of an outer-scale eddy-turnover time, $L/u_L \sim 6$ ns, a period that is comparable to the time that
has elapsed between the initial flow collision and the magnetic field measurements.  That the magnetized plasma is in a saturated state
is corroborated by the FLASH simulation results (see Figure 13b in the Supplementary Information).   

If saturation is reached, the magnetic field energy should become comparable to the turbulent kinetic energy at the outer scale.
We find $ B_{rms}^2/\mu_0 \rho u_L^2 \approx 0.04$ (where $\rho$ is the plasma mass density and we have taken
$B_{rms} \approx 120$ kG). Because the field distribution is expected to be quite intermittent and because Rm in our experiment
is unlikely to be asymptotically large compared to the dynamo threshold value, it is reasonable
that the mean magnetic energy density is quantitatively smaller than the kinetic energy density
\cite{Meneguzzi1981HelicalDynamos,Schekochihin2004,Beresnyak2012}. 
However, a good indication that the magnetic field has reached a dynamically saturated state is that it is
dynamically strong in the most intense structures, which are not necessarily volume filling.
To find an upper experimental bound on the maximum field, $B_{max}$, we assume that the deflections acquired by the imaging protons
across the plasma come from an interaction with a single structure. The strongest individual structure in the reconstructed path-integrated
image has scale $\ell_B \sim 140 \, \mu \mathrm{m}$ with a path-integrated field of 6 kG cm. This gives $B_{max} \lesssim 430 \, \mathrm{kG}$,
which leads to $B_{max}^2/(\mu_0 \rho u_L^2) \lesssim 0.5$, consistent with dynamical strength.

Our results appear to provide a consistent picture of magnetic field amplification by turbulent motions, in agreement with the longstanding theoretical expectation
that turbulent dynamo is the dominant process in achieving dynamical equipartition between kinetic and magnetic energies in high magnetic Reynolds number plasmas 
found in many astrophysical environments.

\acknow{The research leading to these results has received funding from the European Research Council under the European Community's Seventh Framework Programme (FP7/2007-2013) / ERC grant agreements no. 256973 and 247039, the U.S. Department of Energy under Contract No. B591485 to Lawrence Livermore National Laboratory, Field Work Proposal No. 57789 to Argonne National Laboratory, and grants no. DE-NA0002724 and DE-SC0016566 to the University of Chicago. We acknowledge support from the National Science Foundation under grant PHY-1619573. This work was also supported in part by NIH through resources provided by the Computation Institute and the Biological Sciences Division of the University of Chicago and Argonne National Laboratory, under grant S10 RR029030-01. Awards of computer time were provided by the U.S. Department of Energy Innovative and Novel Computational Impact on Theory and Experiment (INCITE) and ASCR Leadership Computing Challenge (ALCC) programs. This research used resources of the Argonne Leadership Computing Facility at Argonne National Laboratory, which is supported by the Office of Science of the U.S. Department of Energy under contract DE-AC02-06CH11357. We acknowledge funding from grants 2007-0093860 and 2016R1A5A1013277 of the National Research Foundation of Korea. Support from AWE plc., the Engineering and Physical Sciences Research Council (grant numbers EP/M022331/1 and EP/N014472/1) and the Science and Technology Facilities Council of the United Kingdom is acknowledged.}

\showacknow

\bibliography{Mendeley.bib,Add_Ref.bib}

\begin{thebibliography}{10}

\bibitem{Carilli2002}
Carilli CL, Taylor GB (2002) {Cluster Magnetic Fields}.
\newblock {\em Annual Review of Astronomy and Astrophysics} 40:319.

\bibitem{Subramanian2006a}
Subramanian K, Shukurov A, Haugen NEL (2006) {Evolving turbulence and magnetic
  fields in galaxy clusters}.
\newblock {\em Monthly Notices of the Royal Astronomical Society}
  366(4):1437--1454.

\bibitem{Ryu2008}
Ryu D, Kang H, Cho J, Das S (2008) {Turbulence and Magnetic Fields in the
  Large-Scale Structure of the Universe}.
\newblock {\em Science} 320(5878):909--912.

\bibitem{Miniati2015}
Miniati F, Beresnyak A (2015) {Self-similar energetics in large clusters of
  galaxies}.
\newblock {\em Nature} 523(7558):59--62.

\bibitem{Zweibel1997}
Zweibel EG, Heiles C (1997) {Magnetic fields in galaxies and beyond}.
\newblock {\em Nature} 385(6):131--136.

\bibitem{Schekochihin2006}
Schekochihin aa, Cowley SC (2006) {Turbulence, magnetic fields, and plasma
  physics in clusters of galaxies}.
\newblock {\em Physics of Plasmas} 13(5):056501.

\bibitem{Batchelor1950}
Batchelor GK (1950) {On the Spontaneous Magnetic Field in a Conducting Liquid
  in Turbulent Motion} in {\em Proceedings of the Royal Society of London.
  Series A}.
\newblock Vol.{} 201, pp. 405--416.

\bibitem{Biermann1951}
Biermann L, Schl{\"{u}}ter A (1951) {Cosmic Radiation and Cosmic Magnetic
  Fields. II. Origin of Cosmic Magnetic Fields}.
\newblock {\em Physical Review} 29(29).

\bibitem{Schekochihin2004}
{Schekochihin} AA, {Cowley} SC, {Taylor} SF, {Maron} JL, {McWilliams} JC (2004)
  {Simulations of the Small-Scale Turbulent Dynamo}.
\newblock {\em Astrophys. J.} 612:276--307.

\bibitem{Beresnyak2012}
Beresnyak A (2012) {Universal Nonlinear Small-Scale Dynamo}.
\newblock {\em Physical Review Letters} 108:035002.

\bibitem{Haugen2004}
{Haugen} NE, {Brandenburg} A, {Dobler} W (2004) {Simulations of nonhelical
  hydromagnetic turbulence}.
\newblock {\em Phys. Rev. E} 70(1):016308.

\bibitem{Kulsrud1997}
Kulsrud RM, Cen R, Ostriker JP, Ryu D (1997) {The protogalactic origin for
  cosmic magnetic fields}.
\newblock {\em Astrophysical Journal} 480:481--491.

\bibitem{Stamper1971}
Stamper JA et~al. (1971) {Spontaneous Magentic Fields in Laser-Produced
  Plasmas}.
\newblock {\em Physical Review Letters} 26(17):1012.

\bibitem{Gregori2012}
Gregori G et~al. (2012) {Generation of scaled protogalactic seed magnetic
  fields in laser-produced shock waves}.
\newblock {\em Nature} 481(7382):480--483.

\bibitem{Meneguzzi1981HelicalDynamos}
Meneguzzi M, Frisch U, Pouquet A (1981) {Helical and Nonhelical Turbulent
  Dynamos}.
\newblock {\em Physical Review Letters} 47(15):1060--1064.

\bibitem{Schekochihin2007}
Schekochihin aa et~al. (2007) {Fluctuation dynamo and turbulent induction at
  low magnetic Prandtl numbers}.
\newblock {\em New Journal of Physics} 9(8):300.

\bibitem{Kulsrud1995ImportantAstrophysics}
Kulsrud RM (1995) {Important plasma problems in astrophysics}.
\newblock {\em Physics of Plasmas} 2(5):1735.

\bibitem{Gregori2015}
Gregori G, Reville B, Miniati F (2015) {The generation and amplification of
  intergalactic magnetic fields in analogue laboratory experiments with high
  power lasers}.
\newblock {\em Physics Reports} 601:1--34.

\bibitem{Monchaux2007}
Monchaux R et~al. (2007) {Generation of a Magnetic Field by Dynamo Action in a
  Turbulent Flow of Liquid Sodium}.
\newblock {\em Physical Review Letters} 98(4):044502.

\bibitem{Meinecke2014a}
Meinecke J, Doyle H, Miniati F, Bell A (2014) {Turbulent amplification of
  magnetic fields in laboratory laser-produced shock waves}.
\newblock {\em Nature Physics} 10(June):520--524.

\bibitem{Meinecke2015a}
Meinecke J et~al. (2015) {Developed turbulence and nonlinear amplification of
  magnetic fields in laboratory and astrophysical plasmas}.
\newblock {\em Proceedings of the National Academy of Sciences}
  112(27):8211--8215.

\bibitem{Boehly1997}
Boehly T et~al. (1997) {Initial performance results of the OMEGA laser system}.
\newblock {\em Optics Communications} 133(1-6):495--506.

\bibitem{Tzeferacos2015a}
Tzeferacos P et~al. (2015) {FLASH MHD simulations of experiments that study
  shock-generated magnetic fields}.
\newblock {\em High Energy Density Physics} 17:24--31.

\bibitem{Tzeferacos2016}
{Tzeferacos} P et~al. (2017) Numerical modeling of laser-driven experiments
  aiming to demonstrate magnetic field amplification via turbulent dynamo.
\newblock {\em ArXiv e-prints}.

\bibitem{Zhuravleva2014}
Zhuravleva I et~al. (2014) {the Relation Between Gas Density and Velocity Power
  Spectra in Galaxy Clusters: Qualitative Treatment and Cosmological
  Simulations}.
\newblock {\em The Astrophysical Journal} 788(1):L13.

\bibitem{Evans1969}
Evans DE, Katzenstein J (1969) {Laser light scattering in laboratory plasmas }.
\newblock {\em Reports On Progress In Physics} 32:207.

\bibitem{Inogamov2003}
Inogamov NA, Sunyaev RA (2003) {Turbulence in Clusters of Galaxies and X-ray
  Line Pro fi les}.
\newblock {\em Astronomy Letters} 29(12):791--824.

\bibitem{Li2006}
Li C et~al. (2006) {Measuring E and B Fields in Laser-Produced Plasmas with
  Monoenergetic Proton Radiography}.
\newblock {\em Physical Review Letters} 97(13):3--6.

\bibitem{Kugland2012a}
Kugland NL, Ryutov DD, Plechaty C, Ross JS, Park HS (2012) {Invited Article:
  Relation between electric and magnetic field structures and their proton-beam
  images}.
\newblock {\em Review of Scientific Instruments} 83(10):101301.

\bibitem{Graziani2016}
{Graziani} C, {Tzeferacos} P, {Lamb} DQ, {Li} C (2016) {Inferring Morphology
  and Strength of Magnetic Fields From Proton Radiographs}.
\newblock {\em ArXiv e-prints}.

\end{thebibliography}

\end{document}